# Large Diameter Bulk Crystal Growth and Scintillation Characterization of Thallium-Based Ternary Halide Crystals for Detection and Imaging

Rastgo Hawrami (Xtallized Intelligence, Inc.)

**ABSTRACT**

Scintillators cover an important wide range of applications in detection and imaging. In this paper we are presenting growth and performance results of several advanced large diameter thallium-based ternary halide crystals are presented. Intrinsic crystals of $TlMgCl_3$, $TlCaCl_3$, $TlCaBr_3$, and $TlCa(Cl,Br)_3$, as well as europium-doped $TlCa_2Br_5$ and $TlCa(Cl,Br)_3$ are melt-grown by the Bridgman method. These compounds have high $Z_{eff}$ due to thallium and expected to have high physical densities. The best crystal quality and energy resolution (FWHM) at 662 keV are observed for intrinsic $TlMgCl_3$. The primary decay constants for these compounds are in the range of 0.45 to 0.55 μs. All of these compounds have proportional or linear response to γ-ray above 100 keV.

**INDEX KEYWORDS**
Gamma-ray detectors. Inorganic scintillators. Intrinsic scintillators. High detection efficiency. Ternary compounds. Thallium-based scintillators.

**INTRODUCTION**

Advances in high energy physics depend on the advancement in materials research. Research in scintillators for particle detectors in high energy physics is pushed by a need for new, high-performance particle scintillation detectors with better energy resolutions, higher light yields, high densities, fast decay times, and are radiation hard [1]. Along with these desirable properties, these materials must also be stable, low cost, as well as easy to produce in large volumes and massive quantities.

Some of the most studied radiation detector materials currently available in market are inorganic binary compound scintillators [2]. Ternary compounds that are based on these binary compounds are usually formed by adding metal cations. Initial examples of these ternary compounds are online [2], with many more ternary compounds published in recent years. Many these recent advances in materials study for radiation sensors, particularly the latest discovery of high-density thallium-based inorganic scintillation crystals, is spurred by a necessity to improve isotope identification capability, for example, for homeland security applications. These new materials also possess many of the required properties for detection materials in high energy physics applications and research. Continuous search for improved scintillation materials for better radiation detection is important, since an ideal scintillator for such applications has yet to be discovered.

Recently high detection efficiency Tl-based scintillation crystals have attracted good attention from worldwide scintillator researchers. These compounds have been investigated and very promising initial results have been published, for example Ce-doped $Tl_2LaCl_5$ (TLC) [3] as well as *intrinsic* (i.e., not doped) $TlMgCl_3$ (TMC) and $TlCaI_3$ (TCI) [4]. These new compounds are of high densities (< 5 g/cm$^3$), bright (light yields between 31,000 and 76,000 ph/MeV for 662 keV photons), fast decay times (36 ns (89%) for TLC; 46 ns (9%) for TMC; 62 ns (13%) for TCI), and moderate melting points (between 500 and 700°C). As seen further in the published results, Tl-based scintillators such as the ones previously mentioned have promising properties desirable for high energy physics as well as homeland security applications [3-10].

At high photon energy, when pair production interactions producing electron-positron pairs dominates, material properties such as radiation length $X_o$, nuclear interaction length $\lambda_I$, and Moliere radius $R_M$ characterize the amount of matter is traversed by charged particles produced as well as the spread of the electromagnetic showers for these related interactions. Materials with high Z constituents tend to have lower values of $X_o$, $\lambda_I$, and $R_M$. Tables 1 show that oxide-based scintillators have the overall lowest values of $X_o$, $\lambda_I$, and $R_M$, followed by TMC and TCI two representative (intrinsic) Tl-based ternary compounds, hence these Tl-based scintillators can be presented as candidate materials in high energy physics applications.

Table 1. $X_o$, $\lambda_I$, and $R_M$ values for scintillators listed in Table 1, along with NaI, CsI, BGO, LSO, and PbWO$_4$ for comparison.

| $X_o$ (cm) | Scintillator | $R_M$ (cm) | Scintillator | $\lambda_I$ (cm) | Scintillator |
|---|---|---|---|---|---|
| 0.89 | PbWO$_4$ | 1.98 | PbWO$_4$ | 20.3 | PbWO$_4$ |
| 1.09 | BGO | 2.05 | LSO | 20.6 | LSO |
| 1.14 | LSO | 2.16 | BGO | 22.3 | BGO |
| 1.49 | TlCaI$_3$ | 2.83 | TlMgCl$_3$ | 29.4 | TlMgCl$_3$ |
| 1.64 | TlMgCl$_3$ | 3.05 | TlCaI$_3$ | 32.8 | TlCaI$_3$ |
| 1.86 | CsI | 3.54 | CsI | 38.0 | CsI |
| 2.59 | NaI | 4.09 | NaI | 42.1 | NaI |

Following the results of previously studied compounds, Xtallized Intelligence, Inc. (XI, Inc.) has grown several large (> 16-mm) diameter single bulk crystals of intrinsic and doped thallium-based ternary compound scintillators for high energy physics applications. These compounds include $Tl_aM_bX_c$, where A is metal, such as Mg or Ca, and X are halogens, such as Cl, Br, or I, with a = 1 or 2, b = 1 or 2, and c = 3 or 5. For high energy physics purposes, ideal physical dimensions of a detection material should accommodate at least 20 $X_o$ (e.g., 17.8 cm (7 inches) for PbWO$_4$, 32.8 cm (12.9 inches) for TlMgCl$_3$). Therefore, to obtain amenable crystals, large diameter growth of crack-free, single crystals must be accomplished. In the following sections the growth and crystal analysis of intrinsic crystals of TlMgCl$_3$, TlCaCl$_3$, TlCaBr$_3$, and TlCa(Cl,Br)$_3$, as well as europium-doped TlCa$_2$Br$_5$ and TlCa(Cl,Br)$_3$ are presented.

Table 2. Starting binary compounds and the resulting ternary compounds.

| Starting Binary Compounds | Dopant | Ternary Compounds |
|---|---|---|
| TlCl + MgCl$_2$ | | TlMgCl$_3$ |
| TlCl + CaCl$_2$ | | TlCaCl$_3$ |
| TlCl + CaBr$_2$ | | TlCaBr$_3$ |
| TlCl$_x$ + TlBr$_{1-x}$ + CaCl$_x$ + CaBr$_{2-x}$ | | TlCa(Cl,Br)$_3$ |
| TlCl + 2 CaBr$_2$ | EuBr$_2$ | TlCa$_2$Br$_5$:Eu |
| TlCl$_x$ + TlBr$_{1-x}$ + CaCl$_x$ + CaBr$_{2-x}$ | EuCl$_2$ or EuBr$_2$ | TlCa(Cl,Br)$_3$:Eu |

**EXPERIMENTAL METHODS**

Each of the Tl-based ternary compounds were grown by stoichiometric mixing of the respective binary compounds. Table 2 shows the binary compounds, dopants, and the resulting ternary compounds. Starting materials of the highest available purity were purchased from US-based chemical vendors. For each targeted ternary compound, stoichiometric amounts of the binary compounds and dopants were purified if necessary using different techniques of purifications and loaded into a freshly cleaned and baked quartz ampoule. Material loading was conducted inside an inert glove box with an argon atmosphere. Using a turbo vacuum pump, the loaded ampoule was dehydrated, if necessary, and then sealed in a high vacuum. The ampoule was placed in a two-zone vertical furnace for growth with the Bridgman method. The top zone of the furnace was set a few degrees above the melting temperature of the compound to ensure a complete melt. The bottom zone was set a few degrees below the crystallization temperature for the compound. The growth was conducted at a rate of 15-25 mm/day. After the crystallization was completed, the ampoule was cooled down 75-150$^{o}$C/day to room temperature. The crystal was then retrieved from the growth ampoule and samples were extracted by slicing the boule with a diamond wire saw. Lapping and polishing were conducted using $Al_2O_3$ and/or SiC polishing pads with mineral oil as lubricant.

Sample characterization was done by encapsulating a sample in a hermetic packaging or by immersing a sample in mineral oil contained in a quartz cup surrounded with diffuse reflectors (Teflon tape and Gore ® pad) for a quick test. For the latter measurement configuration, the quartz cup was coupled to a Hamamatsu R6231-100 photomultiplier tube (PMT) with BC-630 silicone optical coupling compound. Gamma-ray spectroscopy with $^{137}$Cs was done to determine the energy resolution of each scintillator at 662 keV. Other gamma ray sources, such as $^{241}$Am, $^{133}$Ba, $^{22}$Na, and $^{60}$Co were used to determine both energy resolutions and relative light yield (i.e., non-proportionality) data. Decay time was determined by using a CAEN DT5720 digitizer to collect the PMT anode signals, which were then analyzed offline.

## RESULTS AND ANALYSIS
### $TlMgCl_3$

Started with ∅16-mm growths, several good quality transparent ∅1″ $TlMgCl_3$ crystal boules were then grown at XI, Inc., two of which are shown in Fig. 1. Thermal stress that occurred during ampoule cooling produced a crack visible in the first boule, which measured approximately 10 cm in length (Fig. 1a). Single transparent crack free crystal samples of $TlMgCl_3$, however, were successfully obtained from this boule (Fig. 1a). With better quality purified materials, the second boule, which measures approximately 18 cm was grown single, transparent crack free. Mechanical stress due to improper crystal processing produced a chip visible in the obtained sample (Fig. 1b). Regardless of these aforementioned imperfections, growth of ∅1″ TlMgCl$_3$ single crystal boules with sizeable lengths is achievable.

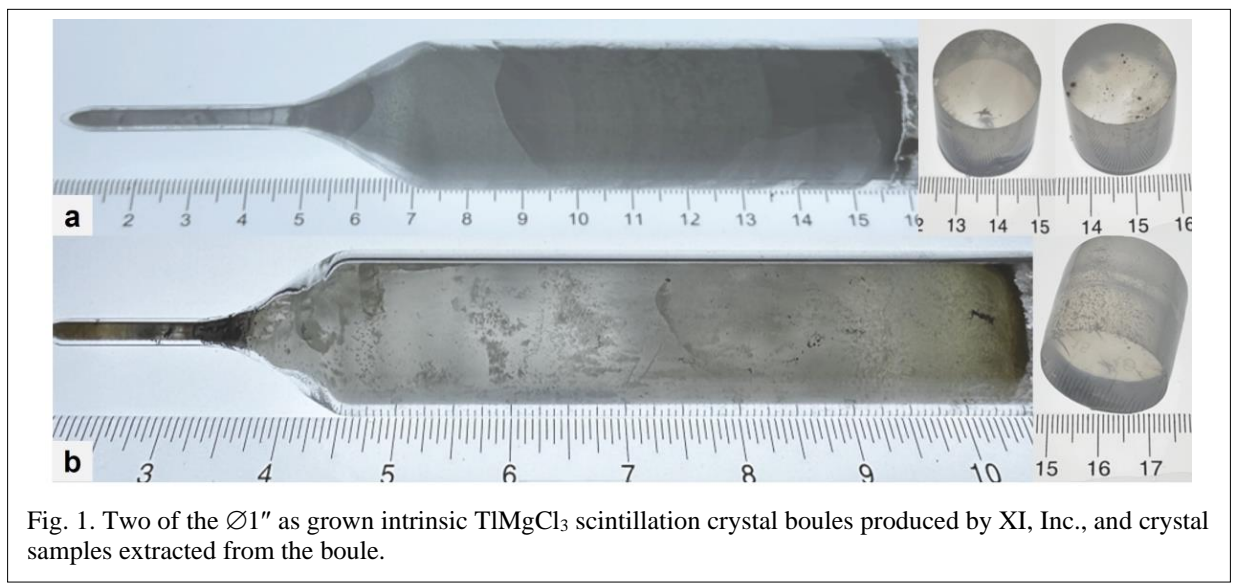

Fig. 1. Two of the ∅1″ as grown intrinsic $TlMgCl_3$ scintillation crystal boules produced by XI, Inc., and crystal samples extracted from the boule.

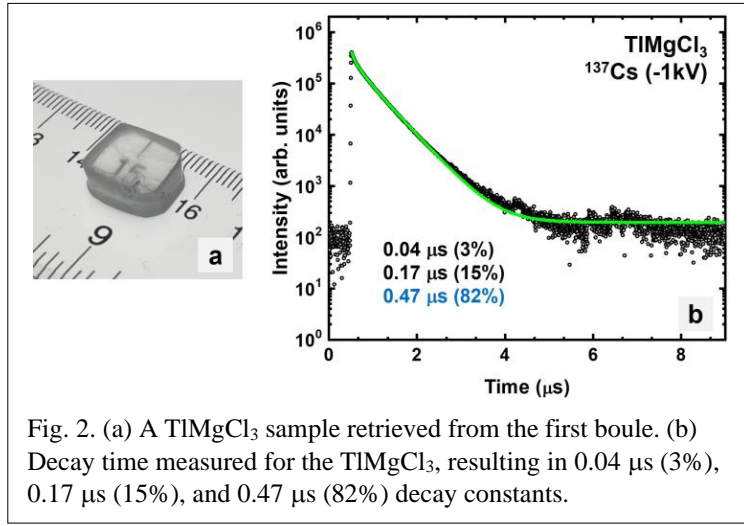

Fig. 2. (a) A TlMgCl$_3$ sample retrieved from the first boule. (b) Decay time measured for the TlMgCl$_3$, resulting in 0.04 μs (3%), 0.17 μs (15%), and 0.47 μs (82%) decay constants.

Fig. 2a shows one of the polished TlMgCl$_3$ crystal samples from the first boule. The temporal profile of the 5×7×10 mm$^3$ sample (Fig. 2b) was analyzed for decay time with three exponential decay functions, resulting in decay constants of 0.04 μs (3%), 0.17 μs (15%), and 0.47 μs (82%).

$^{137}$Cs samples collected by other thin samples collected from along the growth direction of the first boule show a similar detection performance (energy resolution ~ 4.6% at FWHM for 662 keV), indicating that the boule was uniform (Fig. 3). Uniformity in the detector volume is important especially when an application requires that the entire detector volume be involved in detection.

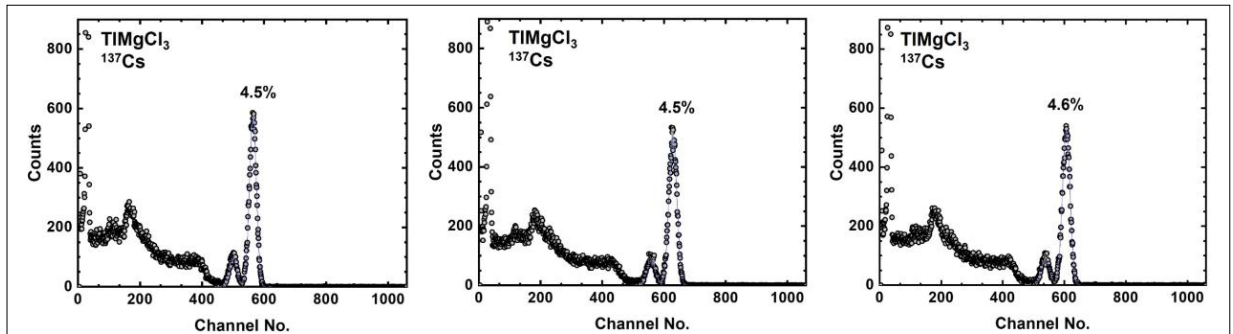

Fig. 3. $^{137}$Cs spectra collected by samples extracted from the first TlMgCl$_3$ boule along the growth direction. The similarity of performance indicates a uniformly grown boule.

A thicker TlMgCl$_3$ crystal (12×15×15 mm$^3$), also obtained from the first boule, collected $^{137}$Cs spectra with different amplifier shaping time (Fig. 4a). Energy resolution of 4.4% (FWHM) at 662 keV was obtained for 4 μs shaping time. The non-proportionality data in Fig. 4b also shows that the response of TlMgCl$_3$ to γ-rays is linear for energy above 30 keV. The decay constants measured from the temporal profile in Fig. 4c are similar to the thinner TlMgCl$_3$ sample.

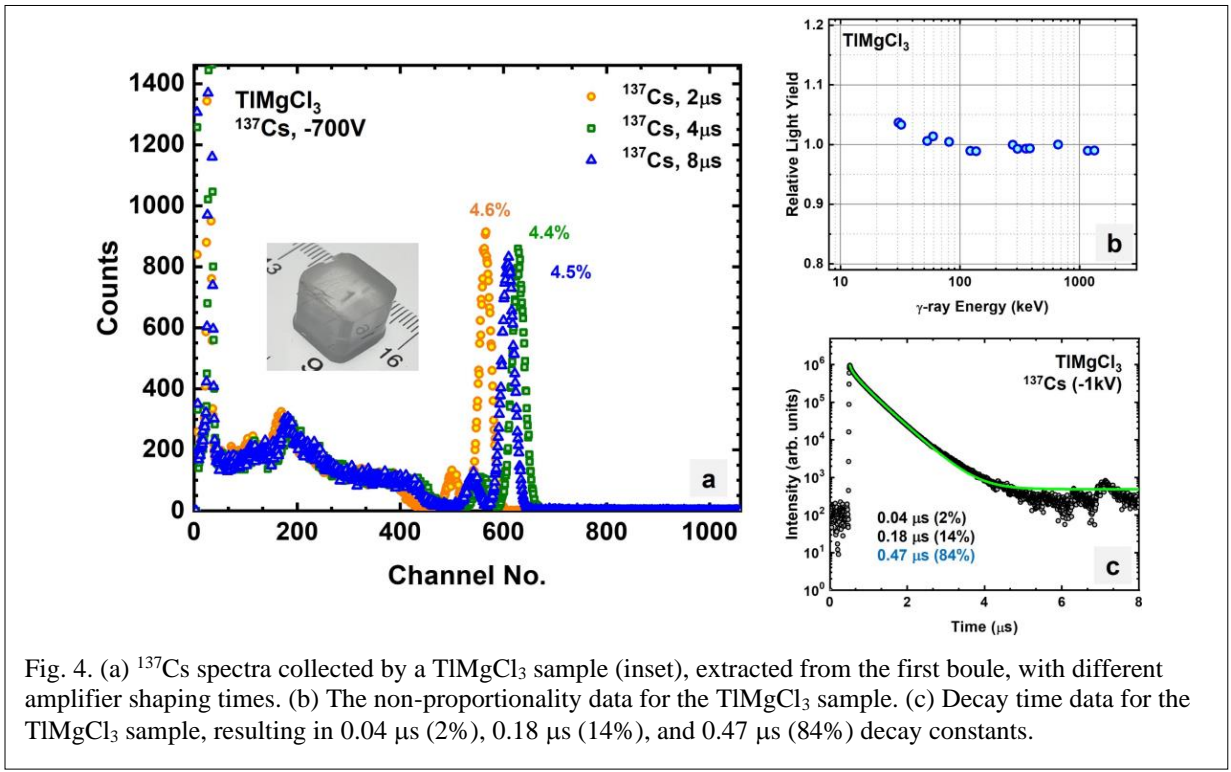

Fig. 4. (a) $^{137}$Cs spectra collected by a TlMgCl$_3$ sample (inset), extracted from the first boule, with different amplifier shaping times. (b) The non-proportionality data for the TlMgCl$_3$ sample. (c) Decay time data for the TlMgCl$_3$ sample, resulting in 0.04 μs (2%), 0.18 μs (14%), and 0.47 μs (84%) decay constants.

A bulk ⌀1″×1″ TlMgCl$_3$ crystal collected $^{137}$Cs spectra with different amplifier shaping times, resulting in a best energy resolution of 3.8% (FWHM) at 662 keV (Fig. 5a). This measurement was slightly better than the results obtained for the smaller crystals, which was probably due to better processing or the shape of the crystal (right cylinder vs. cuboid) that promotes better light collection, better processing (better polishing and no cracking or cleaving), and/or better crystal quality (lack of defects or imperfections). Further study on light collection in TlMgCl$_3$ may be needed. Similar decay time constants to the previous measurements for the smaller crystals were obtained when its temporal profile data was analyzed (Fig. 5b)TlCaCl$_3$

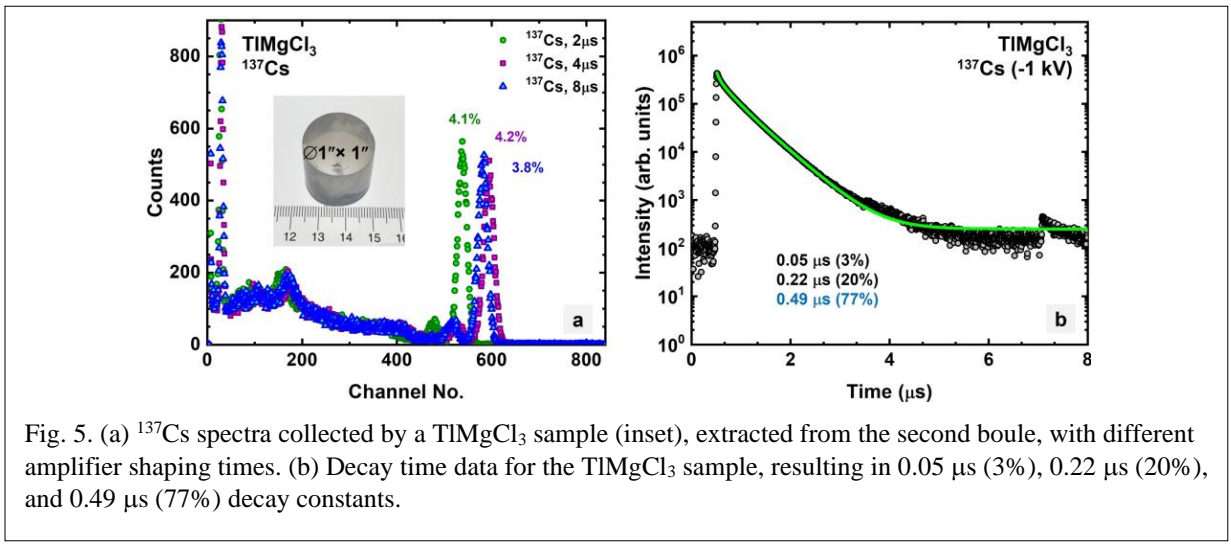

Fig. 5. (a) $^{137}$Cs spectra collected by a TlMgCl$_3$ sample (inset), extracted from the second boule, with different amplifier shaping times. (b) Decay time data for the TlMgCl$_3$ sample, resulting in 0.05 μs (3%), 0.22 μs (20%), and 0.49 μs (77%) decay constants.

Small (16 mm) diameter

A section of an initial as grown ⌀16-mm TlCaCl$_3$ crystal boule, using as received chemicals from vendors without purification, is shown in Fig. 6a. One of the roughly polished samples that were extracted by slicing the boule with a diamond wire saw and then later shaped into a rectangular cuboid (12×12×10 mm$^3$) is shown in Fig. 6b.

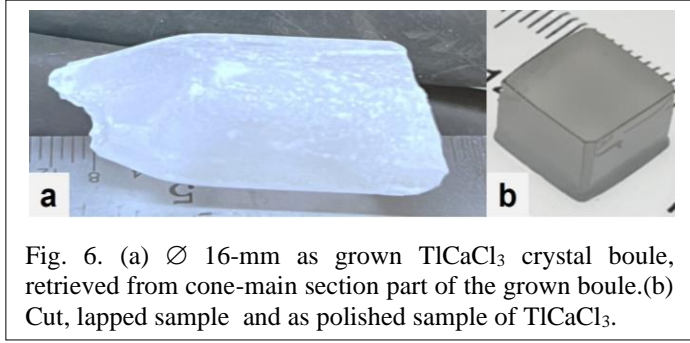

$^{137}$Cs spectra collected with the TlCaCl$_3$ crystal, with different amplifier shaping times, resulted in an energy resolution of 5.2% (FWHM) at 662 keV (Fig. 7a). Non-proportionality data in Fig. 7b shows that the response of TlCaCl$_3$ to γ-rays is linear for energy above 30 keV. Analysis on the temporal profile in Fig. 7c shows that the decay constants for TlCaCl$_3$ are 0.03 μs (1%), 0.28 μs (32%), and 0.69 μs (67%).

Fig. 6. (a) ⌀ 16-mm as grown TlCaCl$_3$ crystal boule, retrieved from cone-main section part of the grown boule. (b) Cut, lapped sample and as polished sample of TlCaCl$_3$.

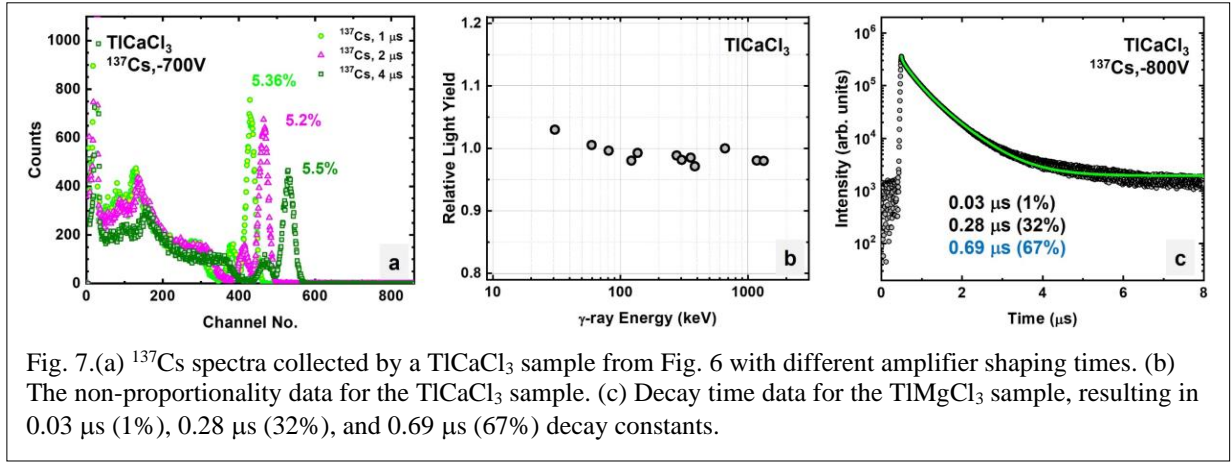

Fig. 7.(a) $^{137}$Cs spectra collected by a TlCaCl$_3$ sample from Fig. 6 with different amplifier shaping times. (b) The non-proportionality data for the TlCaCl$_3$ sample. (c) Decay time data for the TlMgCl$_3$ sample, resulting in 0.03 μs (1%), 0.28 μs (32%), and 0.69 μs (67%) decay constants.

Fig. 8 shows the measurement results ($^{137}$Cs spectra and decay times measurement) of several polished TlCaCl$_3$ samples that were retrieved along the direction of the growth. The $^{137}$Cs spectra shows similar energy resolution (approximately 4.6% FWHM at 662 keV) and similar full energy peak positions, which indicates that the boule was uniform in quality and performance.

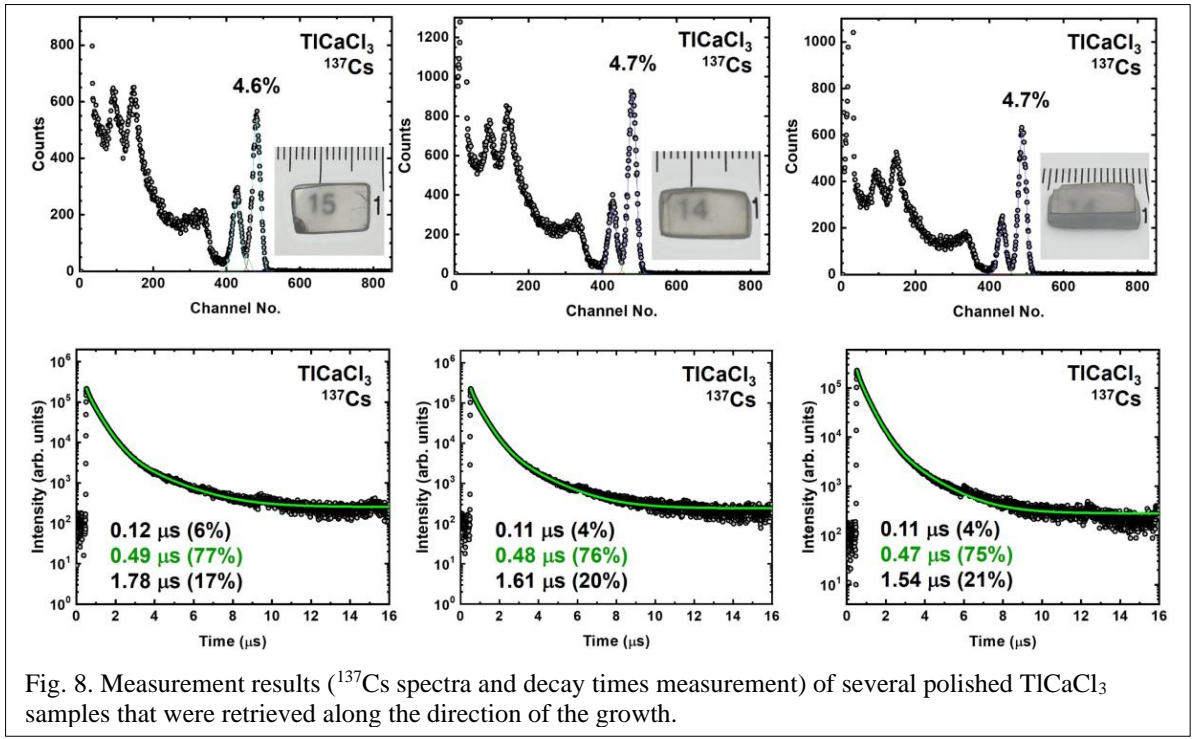

Fig. 8. Measurement results ($^{137}$Cs spectra and decay times measurement) of several polished TlCaCl$_3$ samples that were retrieved along the direction of the growth.

One-inch diameter

An as grown bulk ∅1″ TlCaCl$_3$ single crack free crystal boule with an approximate length of 9.5 cm is shown in Fig. 9a. The outside of the boule was covered with a thin translucent layer that may have been organics or impurities due to excess of chloride reacting with the quartz ampoule. Under this translucent layer was a single crack free crystal, a sample of which was shown in Fig 9b before packaging. A sample cut from the boule was shaped into a rectangular cuboid (Fig. 9b) with the size of 16×16×25 mm$^3$, which was subsequently polished and hermetically encapsulated (Fig. 9c). The encapsulated crystal was characterized by collecting $^{137}$Cs and $^{152}$Eu spectra, as well as collecting the PMT anode signals to determine the temporal profile and determine decay time constants.

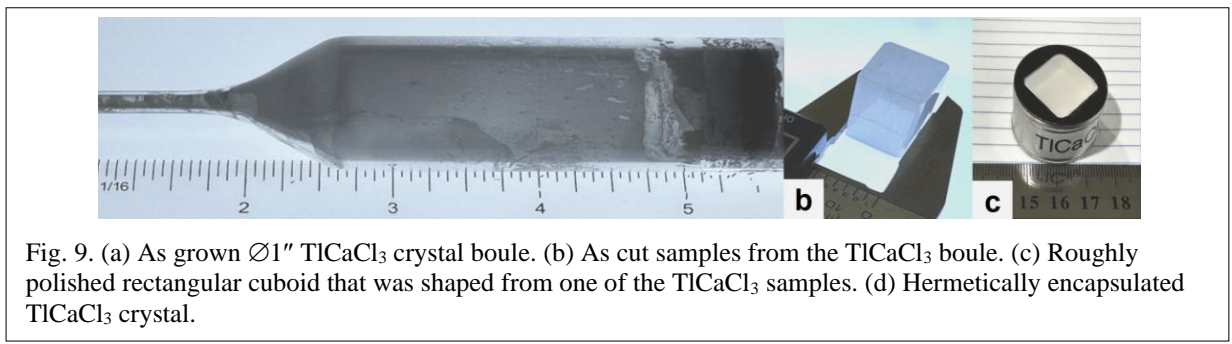

Fig. 9. (a) As grown ∅1″ TlCaCl$_3$ crystal boule. (b) As cut samples from the TlCaCl$_3$ boule. (c) Roughly polished rectangular cuboid that was shaped from one of the TlCaCl$_3$ samples. (d) Hermetically encapsulated TlCaCl$_3$ crystal.

Fig. 10a shows the $^{137}$Cs spectrum collected by the encapsulated TlCaCl$_3$ sample with measured 4.6% (FWHM) energy resolution at 662 keV. Decay constants of 0.1 μs (3%), 0.46 μs (72%), and 1.28 μs (25%) were measured (Fig. 10b). $^{152}$Eu spectrum was also collected (Fig. 10c), from which the relative light yield data were calculated to determine the non-proportionality behavior of TlCaCl$_3$ (Fig. 10d). As in the case of the smaller sized TlCaCl$_3$ crystals, the crystal had a linear response to γ-rays with energy above 100 keV.

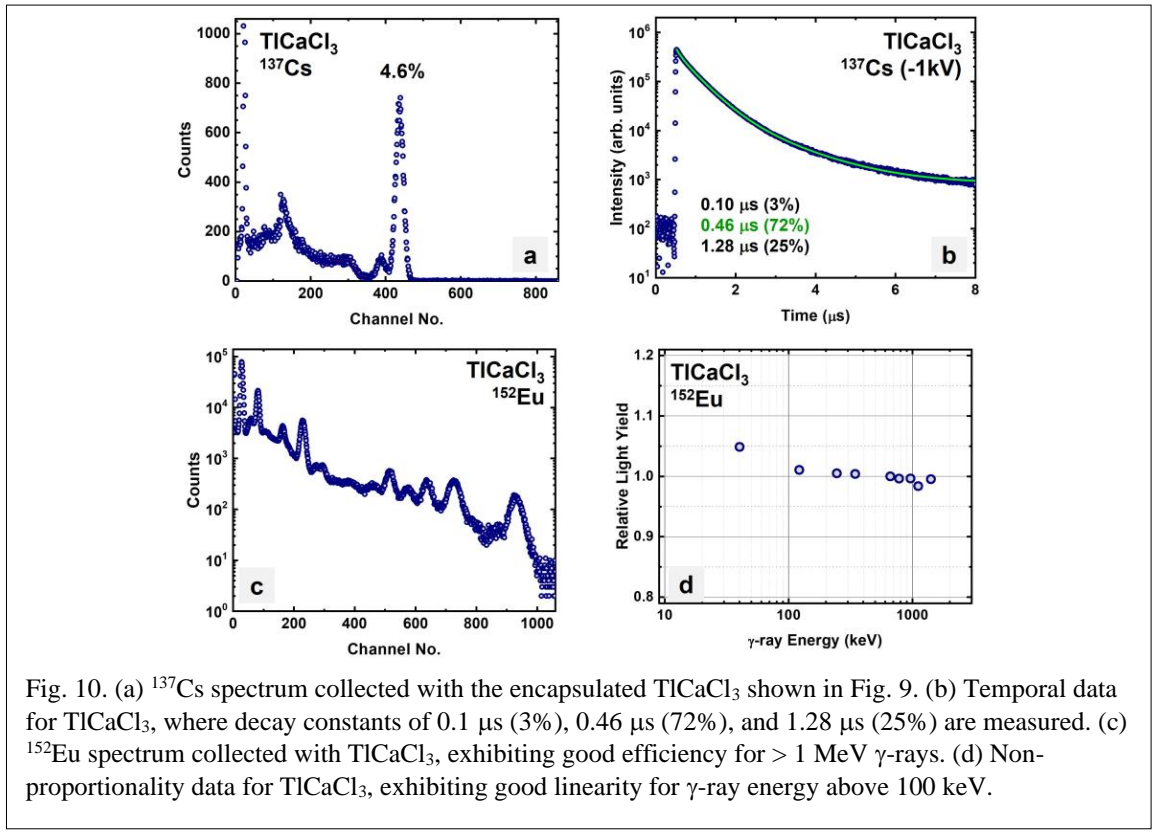

Fig. 10. (a) $^{137}$Cs spectrum collected with the encapsulated TlCaCl$_3$ shown in Fig. 9. (b) Temporal data for TlCaCl$_3$, where decay constants of 0.1 μs (3%), 0.46 μs (72%), and 1.28 μs (25%) are measured. (c) $^{152}$Eu spectrum collected with TlCaCl$_3$, exhibiting good efficiency for > 1 MeV γ-rays. (d) Non-proportionality data for TlCaCl$_3$, exhibiting good linearity for γ-ray energy above 100 keV.

### TlCaBr$_3$

Fig. 11a shows an as grown ⌀14-mm TlCaBr$_3$ crystal boule with an approximate length of 7.5 cm. The visible crack in the mid-section of the boule was due to a power failure during the cooling process. $^{137}$Cs spectra collected with different amplifier shaping times by a TlCaBr$_3$ sample are shown in Fig. 11b, with measured energy resolution of 5.3% (FWHM) at 662 keV. Comparison of $^{137}$Cs spectra collected with a TlCaBr$_3$ sample from the ⌀13×20-mm boule and a ⌀1″×1″ NaI:Tl is shown in Fig. 11c, showing that even with a smaller diameter, TlCaBr$_3$ exhibited a larger peak-to-Compton ratio than NaI:Tl. $^{152}$Eu spectrum was also collected (Fig. 11d), from which the relative light yield data were calculated to determine the non-proportionality behavior of TlCaBr$_3$ (Fig. 11e). As in the case of other ternary Tl-based halide crystals described so far, the crystal had a linear response to γ-rays with energy above 100 keV.

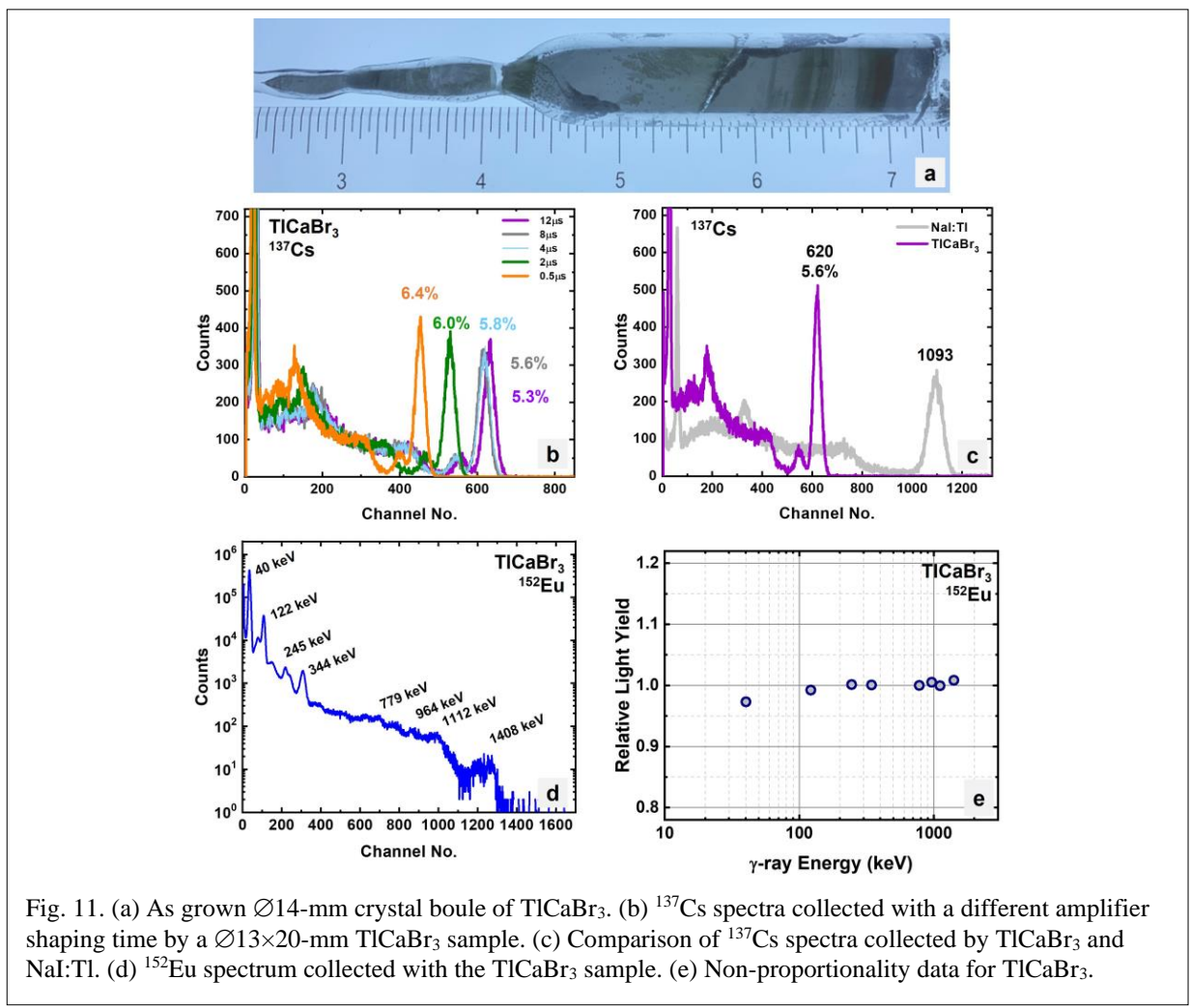

Fig. 11. (a) As grown ⌀14-mm crystal boule of TlCaBr$_3$. (b) $^{137}$Cs spectra collected with a different amplifier shaping time by a ⌀13×20-mm TlCaBr$_3$ sample. (c) Comparison of $^{137}$Cs spectra collected by TlCaBr$_3$ and NaI:Tl. (d) $^{152}$Eu spectrum collected with the TlCaBr$_3$ sample. (e) Non-proportionality data for TlCaBr$_3$.

*Eu-doped TlCa$_2$Br$_5$*

Two as grown ⌀16-mm crystal boules of TlCa$_2$Br$_5$ (top: undoped, bottom: europium-doped) are shown in Fig. 12. The appearance of the crystals were not clear as one can observe from a processed thin sample of Eu-doped TlCa$_2$Br$_5$ (inset picture in Fig. 13a), from which clearly there were more than one phases grown. A dopant (Eu$^{2+}$) was used in attempt to improve scintillation properties like light yield, energy resolution, and decay time.

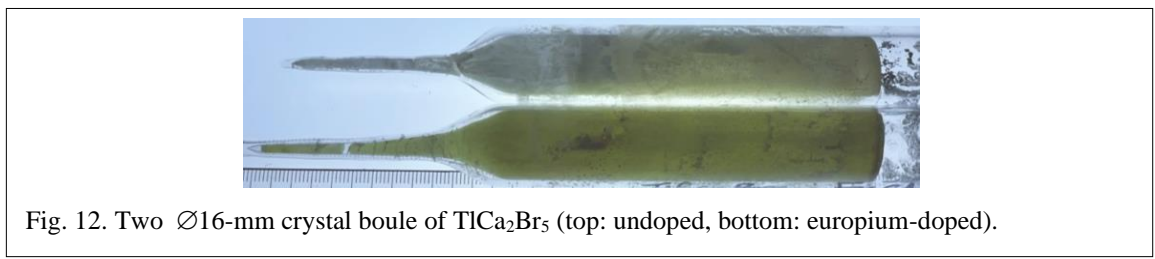

Fig. 12. Two ⌀16-mm crystal boule of TlCa$_2$Br$_5$ (top: undoped, bottom: europium-doped).

A sample extracted from TlCa$_2$Br$_5$:Eu boule was characterized (Fig 13a inset). The crystal was not transparent and appeared to have more than one phases. Fig. 13a shows $^{137}$Cs spectrum collected by the Eu-doped TlCa$_2$Br$_5$ crystal. Energy resolutions of 4.2% (FWHM at 662 keV) was measured. Three decay constants were calculated (Fig. 13(b)): 0.54 μs (77%) and 1.8 μs (23%). According to the relative light yield vs. photon energy (non-proportionality) data (Fig. 13c), TlCa$_2$Br$_5$:Eu has a relatively linear response (± 0.05%) for a wide range of photon energy, especially for energy above 100 keV. Europium as a dopant appears to improve energy resolution and slightly improve light yield. However, crystal quality appears to have been compromised.

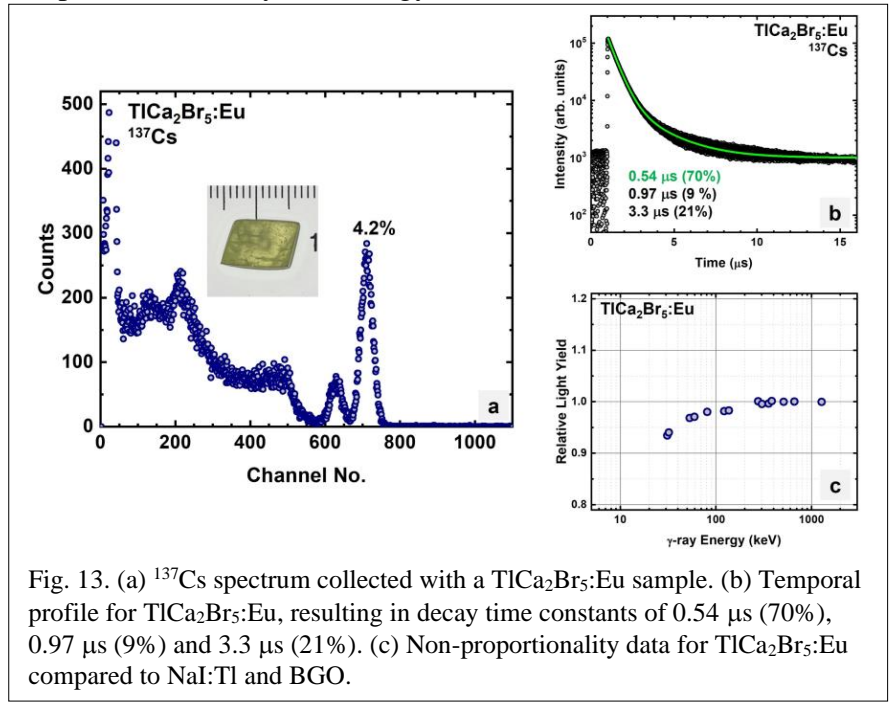

Fig. 13. (a) $^{137}$Cs spectrum collected with a TlCa$_2$Br$_5$:Eu sample. (b) Temporal profile for TlCa$_2$Br$_5$:Eu, resulting in decay time constants of 0.54 μs (70%), 0.97 μs (9%) and 3.3 μs (21%). (c) Non-proportionality data for TlCa$_2$Br$_5$:Eu compared to NaI:Tl and BGO.

*Mixed halide TlCa(Cl,Br)$_3$*

As seen in the previous sections, for the thallium calcium halide system, there are several avenues that were attempted to improve the intrinsic and/or scintillation properties of the crystals. First, undoped TlCaBr$_3$ was grown as the base for comparison. Second, Eu-doped TlCa$_2$Br$_5$ was grown, resulting in slight improvement in light yield and energy resolution, while the crystal quality suffers. Third, (undoped) mixed halide TlCa(Cl,Br)$_3$ was grown. Fig. 14a shows $^{137}$Cs spectra collected with different amplifier shaping times by a TlCa(Cl,Br)$_3$ sample.

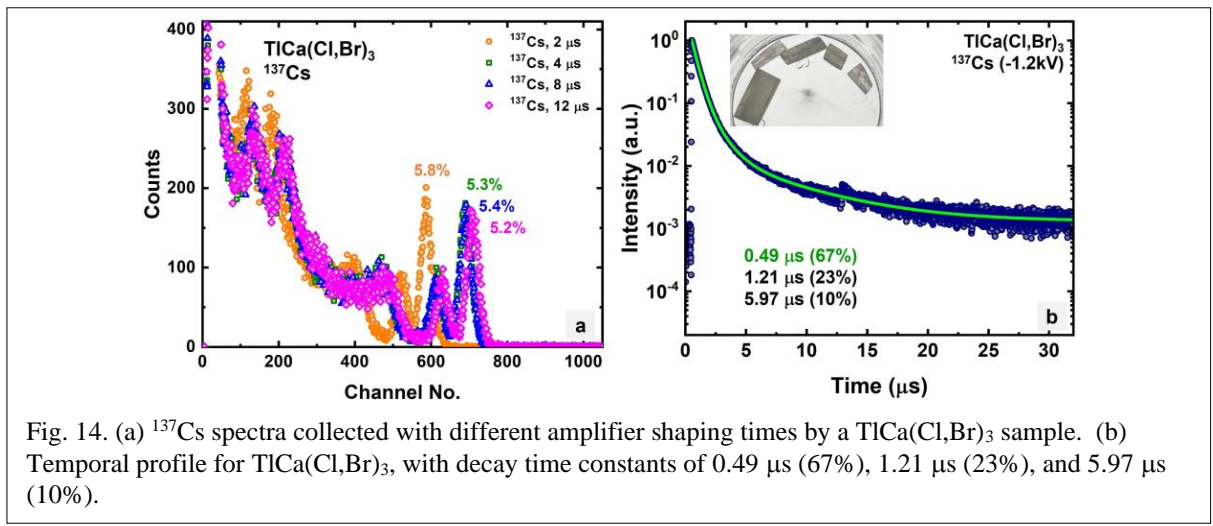

Fig. 14. (a) $^{137}$Cs spectra collected with different amplifier shaping times by a TlCa(Cl,Br)$_3$ sample. (b) Temporal profile for TlCa(Cl,Br)$_3$, with decay time constants of 0.49 μs (67%), 1.21 μs (23%), and 5.97 μs (10%).

Compared to intrinsic TlCaBr$_3$, no improvement was observed in either energy resolution or light yield. Fig. 14b shows the temporal profile for TlCa(Cl,Br)$_3$, with decay constants of 0.49 μs (67%), 1.21 μs (23%), and 5.97 μs (10%).

*Eu-doped TlCa(Cl,Br)$_3$*

Following the study presented in the previous sections, europium doped TlCa(Cl,Br)$_3$ were grown, with the results shown in Fig. 15. The crystals extracted from along the boule were not transparent, however, their performance was similar (energy resolution 5.5-5.8% (FWHM) at 662 keV; primary decay constant 0.50-0.58 μs), indicating that the boule was grown uniformly. There was no observable improvement to the energy resolution nor to the decay constants, as compared to TlCaBr$_3$.

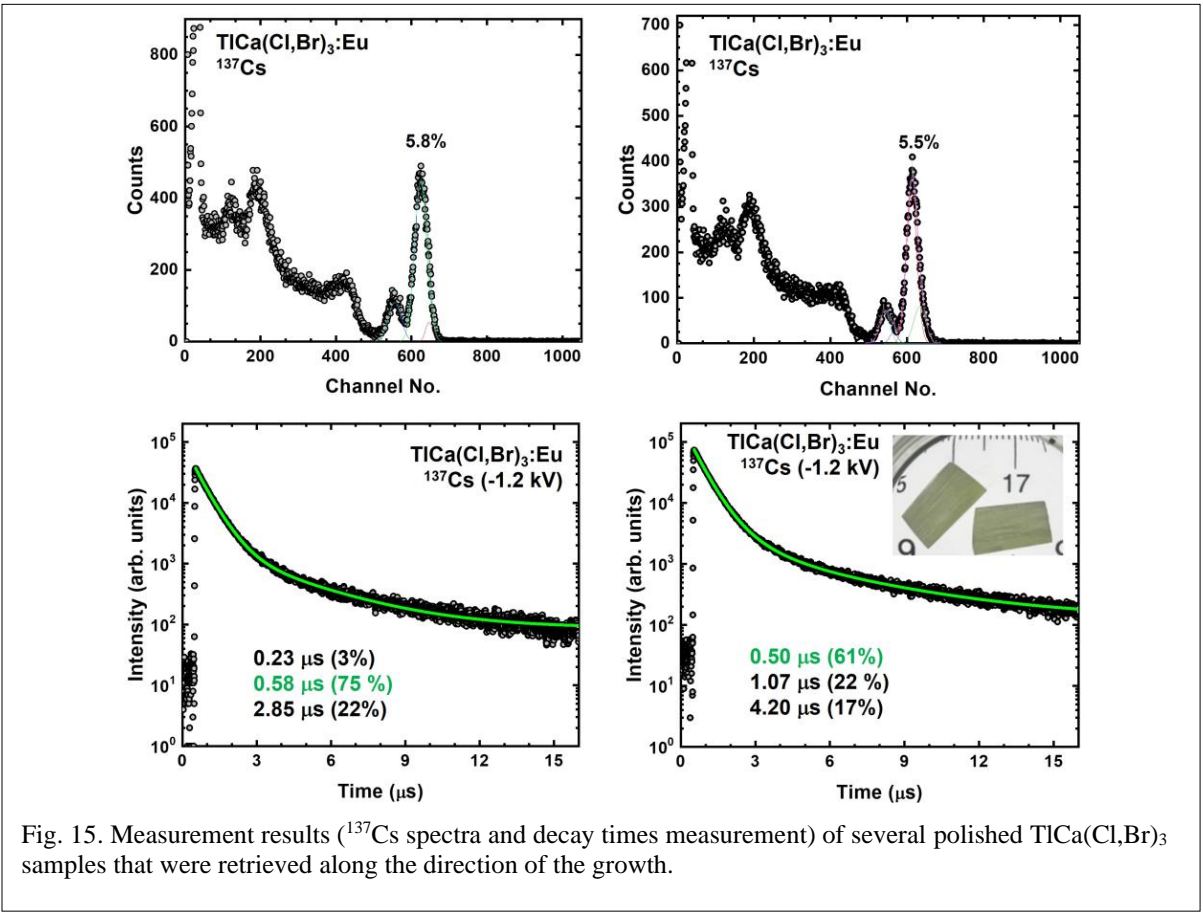

Fig. 15. Measurement results ($^{137}$Cs spectra and decay times measurement) of several polished TlCa(Cl,Br)$_3$ samples that were retrieved along the direction of the growth.

*(Cs,Tl)$_3$Cu$_2$I$_5$*

Initial ⌀1″ crystals of intrinsic Cs$_3$Cu$_2$I$_5$ and Cs$_3$Cu$_2$I$_5$ co-doped with Li$^+$ and Tl$^+$ were grown at XI, Inc. Fig. 16a shows $^{137}$Cs spectra with different shaping times collected by a platelet like sample of intrinsic Cs$_3$Cu$_2$I$_5$ (inset picture), with energy resolutions of 4.8-5.1% (FWHM) at 662 keV were measured. Similar performance was obtained for Cs$_3$Cu$_2$I$_5$ co-doped with Li$^+$ and Tl$^+$ (Fig. 16b).

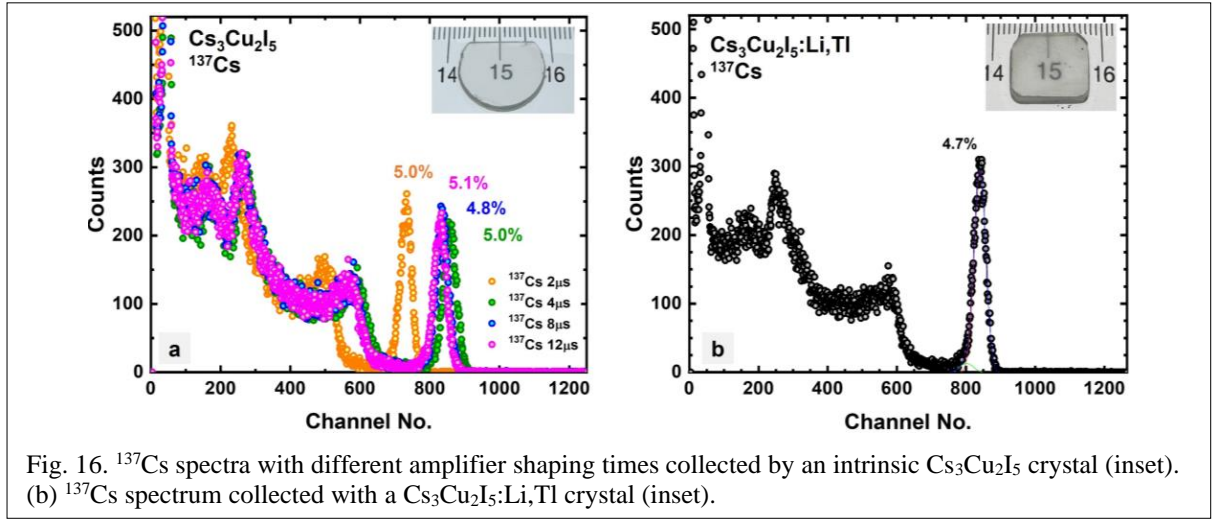

Fig. 16. $^{137}$Cs spectra with different amplifier shaping times collected by an intrinsic Cs$_3$Cu$_2$I$_5$ crystal (inset). (b) $^{137}$Cs spectrum collected with a Cs$_3$Cu$_2$I$_5$:Li,Tl crystal (inset).

## CONCLUSIONS

Table 3 summarizes the results and properties of ternary Tl-halide crystals successfully investigated and grown at XI, Inc. for imaging and high particle physics applications. These compounds have high $Z_{eff}$ due to thallium and expected to have high physical densities. The best crystal quality and energy resolution (FWHM) at 662 keV were observed for intrinsic TlMgCl$_3$. Additionally, TlMgCl$_3$ is not hygroscopic. The primary decay constants for these compounds are in the range of 0.45 to 0.55 µs. All of these compounds have proportional or linear response to γ-ray above 100 keV.

Table 3. Results and properties of ternary Tl-halide crystals grown

| Compound | $Z_{eff}$ | $E_R$ | τ (µs) |
|---|---|---|---|
| TlMgCl$_3$ | 69.7 | 3.8% | 0.47 |
| TlCaCl$_3$ | 68.9 | 4.6% | 0.47 |
| TlCaBr$_3$ | 64.3 | 5.3% | 0.49 |
| TlCa(Cl,Br)$_3$ | 66.2 | 5.2% | 0.49 |
| TlCa$_2$Br$_5$:Eu | 58.9 | 4.2% | 0.52 |
| TlCa(Cl,Br)$_3$:Eu | 64.3 | 5.5% | 0.50 |

## ACKNOWLEDGMENT
This work was supported in part by the U.S. Department of Energy Grant No. DE-SC0022792.